\documentclass[english,reprint]{revtex4-1}
\usepackage[T1]{fontenc}
\usepackage[latin9]{inputenc}
\usepackage{babel}
\usepackage{amsmath}
\usepackage{amssymb}
\usepackage{graphicx}
\usepackage{esint}
\usepackage[unicode=true,pdfusetitle,
 bookmarks=true,bookmarksnumbered=false,bookmarksopen=false,
 breaklinks=false,pdfborder={0 0 1},backref=section,colorlinks=false]
 {hyperref}
\hypersetup{
 pdftex}

\makeatletter
 
 \@ifundefined{textcolor}{}
 {%
   \definecolor{BLACK}{gray}{0}
   \definecolor{WHITE}{gray}{1}
   \definecolor{RED}{rgb}{1,0,0}
   \definecolor{GREEN}{rgb}{0,1,0}
   \definecolor{BLUE}{rgb}{0,0,1}
   \definecolor{CYAN}{cmyk}{1,0,0,0}
   \definecolor{MAGENTA}{cmyk}{0,1,0,0}
   \definecolor{YELLOW}{cmyk}{0,0,1,0}
 }

\makeatother

\begin{document}

\title{Electron spin resonance detected by a superconducting qubit}

\author{Y. Kubo$^{1}$, I. Diniz$^{2}$, C. Grezes$^{1}$, T. Umeda$^{3}$,
J. Isoya$^{3}$, H. Sumiya$^{4}$, T. Yamamoto$^{5}$, H. Abe$^{5}$,
S. Onoda$^{5}$, T. Ohshima$^{5}$, V. Jacques$^{6}$, A. Dréau$^{6}$,
J.-F. Roch$^{6}$, A. Auffeves$^{2}$, D. Vion$^{1}$, D. Esteve$^{1}$,
and P. Bertet$^{1}$}

\affiliation{$^{1}$Quantronics group, SPEC (CNRS URA 2464), CEA-Saclay, 91191
Gif-sur-Yvette, France}

\affiliation{$^{2}$Institut Néel, CNRS, BP 166, 38042 Grenoble, France}

\affiliation{$^{3}$ Research Center for Knowledge Communities, University of
Tsukuba, Tsukuba 305-8550, Japan}

\affiliation{$^{4}$Sumitomo Electric Industries Ltd., Itami 664-001, Japan}

\affiliation{$^{5}$ Japan Atomic Energy Agency, Takasaki 370-1292, Japan}

\affiliation{$^{6}$ LPQM (CNRS UMR 8537), ENS de Cachan, 94235 Cachan, France}

\date{\today}
\begin{abstract}
A new method for detecting the magnetic resonance of electronic spins
at low temperature is demonstrated. It consists in measuring the signal
emitted by the spins with a superconducting qubit that acts as a single-microwave-photon
detector, resulting in an enhanced sensitivity. We implement this
new type of electron-spin resonance spectroscopy using a hybrid quantum
circuit in which a transmon qubit is coupled to a spin ensemble consisting
of NV centers in diamond. With this setup we measure the NV center
absorption spectrum at $30$~mK at an excitation level of $\thicksim15\,\mu_{B}$
out of an ensemble of $10^{11}$ spins. 
\end{abstract}
\maketitle

\section{Introduction}

Electron spin resonance (ESR) spectroscopy at low temperatures is
often complicated by the long spin-lattice energy relaxation time
which can reach minutes at sub-kelvin temperatures. For continuous-wave
(CW) ESR spectroscopy, this implies working at low powers to avoid
saturating the spins; in pulsed ESR this imposes low repetition rates.
In both cases a higher sensitivity for detecting the signal absorbed
or emitted by the spins would be desirable. Recently, tools borrowed
from superconducting quantum electronics have been applied to high-sensitivity
ESR spectroscopy. High-Q superconducting coplanar waveguide resonators
have been used for CW-ESR at millikelvin temperatures with a cryogenic
low-noise HEMT amplifier followed by homodyne detection \cite{Schuster2010,Kubo2010,Majer2011,Bushev2011},
and for pulsed ESR at kelvin temperatures with a commercial ESR spectrometer
\cite{Schuster2012}, yielding promising results in terms of sensitivity.
Here we go one step further and use an on-chip single microwave photon
detector \cite{Chen2011} based on a superconducting qubit \cite{WendinShumeiko2007}
to realize a new type of high-sensitivity low-temperature ESR spectrometer. 

\begin{figure}[!h]
\includegraphics[scale=0.8]{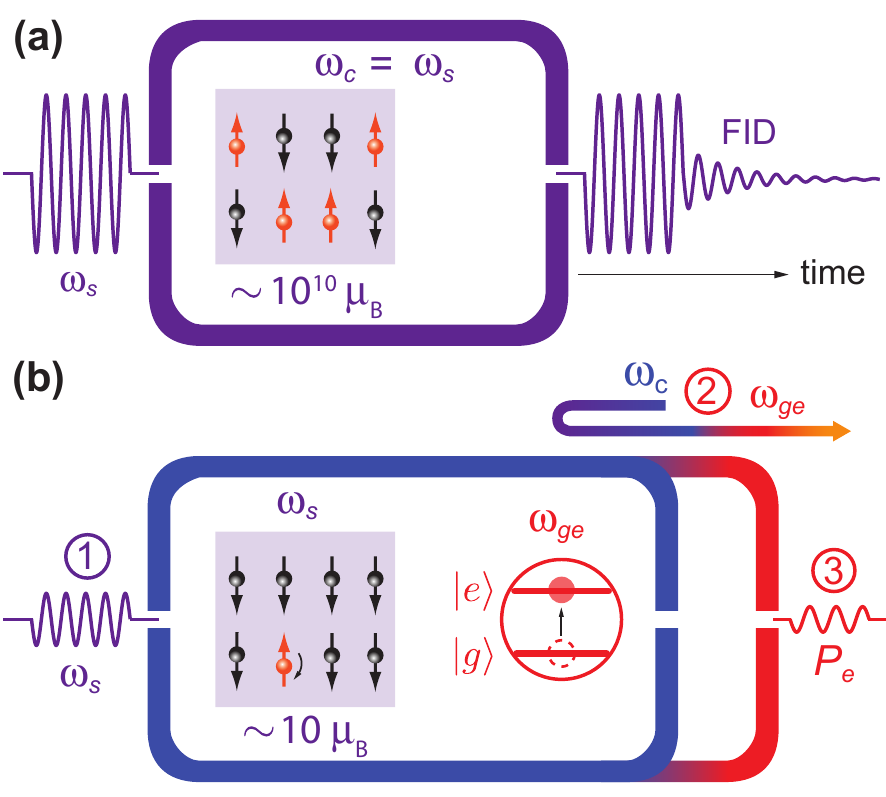}

\caption{Comparison between conventional pulsed electron spin resonance (ESR)
and qubit-detected ESR. \textbf{a}, Conventional ESR: A spin ensemble
is placed in a cavity with frequency $\omega_{c}$ and driven with
a microwave pulse resonant with the cavity. When the ESR frequency
$\omega_{s}$ matches $\omega_{c}$, the spins absorb the microwave
pulse, and emit immediately after a free-induction decay (FID) signal
into the waveguide connected to the cavity. \textbf{b}, Qubit-detected
ESR: The cavity is now frequency-tunable and embeds both the spin
ensemble and a superconducting qubit with frequency $\omega_{ge}$.
In a first step (1), the spins are probed by a spectroscopy pulse
which excites them if its frequency matches $\omega_{s}$. In a second
step (2), the cavity frequency is tuned to $\omega_{c}=\omega_{s}$,
receives the FID signal from the spins, and is afterwards tuned to
transfer this signal to the superconducting qubit at $\omega_{ge}$
. Finally (3) the qubit excited state probability $P_{e}(\omega_{s})$
is measured, mapping the spins absorption spectrum. Very low excitation
powers can be used given the high sensitivity of the method.}

\label{fig1} 
\end{figure}

\begin{figure}[t]
\includegraphics[scale=0.9]{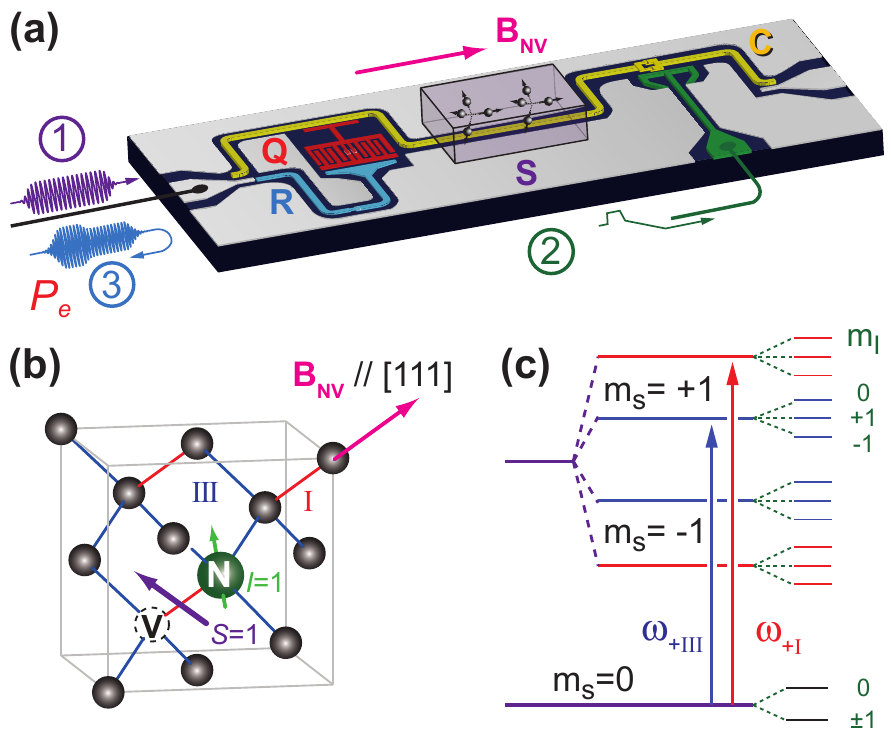}

\caption{\textbf{a,} Sketch of the implementation of qubit-based ESR. The spin
ensemble $S$ consists of NV centers in a diamond crystal. They are
coupled to the frequency-tunable coplanar waveguide resonator $C$
used as the ESR cavity. $C$ also embeds the ESR-detector qubit $Q$,
a superconducting qubit of the transmon type whose state can be readout
with another resonator $R$. Microwave pulses for spin spectroscopy
as well as for qubit readout are sent via an input port coupled both
to $C$ and to $R$. \textbf{b} and \textbf{c}, Sketch and energy
levels of NV centers in diamond. In our setup a DC magnetic field
$\mathbf{B}_{NV}$ is applied along the $[1,1,1]$ direction, resulting
in different Zeeman splittings for centers having the $N-V$ axis
parallel $\mathbf{B}_{NV}$ (ensemble $I$, in red) and those having
their axis along the three other $\left\langle 1,1,1\right\rangle $
axes (ensemble $III$, in blue). The ESR frequencies $\omega_{\pm I,III}$
are further split in three resonance lines due to the hyperfine interaction
with the spin-1 $^{14}N$ nuclear spin \cite{Felton2009}.}

\label{fig2} 
\end{figure}

The principle of our experiment is compared to more conventional ESR
techniques in Fig.~\ref{fig1}. In conventional ESR spectroscopy,
a microwave pulse is applied to an ensemble of spins close to their
resonance frequency $\omega_{s}$ through a low-Q cavity of frequency
$\omega_{c}$ with which the spins are tuned in resonance ($\omega_{s}=\omega_{c}$)
by a magnetic field. After being excited by the pulse, the spins re-radiate
coherently part of the absorbed energy through the cavity into the
detection waveguide, giving rise to a free induction decay (FID) signal
measured by homodyne detection, which yields after Fourier transform
the spin absorption spectrum \cite{ESRnote}. What limits the sensitivity
of a typical commercial CW-spectrometer operating at $300$~K to
$\sim10^{10}\mathrm{spins}/\sqrt{\mathrm{Hz}}$ for a linewidth of
$0.1$~mT and an integration time of $1$~s is the overall noise
temperature of the detection chain. In this work, we replace the detection
chain by a superconducting qubit and its readout circuitry. This results
in an increased sensitivity since a superconducting qubit is a nearly
ideal single microwave photon detector \cite{Chen2011} at its resonance
frequency $\omega_{ge}$. In order to transfer part of the excitation
of the spins to the superconducting qubit, the ESR resonator is made
frequency-tunable and with a high quality factor. Spectroscopy is
performed by first exciting the spins with a weak microwave pulse,
collecting the radiated FID signal with the resonator tuned at $\omega_{s}$,
then transferring this signal to the qubit at $\omega_{ge}$, and
measuring its final state. Repeating this experimental sequence yields
the probability $P_{e}$ to find the qubit in its excited state, which
reproduces the spin absorption spectrum. The sensitivity of such an
ESR spectrometer is set by the efficiency at which signal photons
can be transferred from the spins to the resonator, then to the qubit,
and by the fidelity with which the qubit state can be measured.

\section{Device and experimental setup}

We implement this method using a recently reported hybrid quantum
circuit \cite{KuboNVQubit2011} that includes an ensemble of electronic
spins magnetically coupled to a superconducting resonator, itself
electrically coupled to a superconducting qubit, as sketched in Fig.
\ref{fig2}a. The spins are negatively-charged nitrogen-vacancy (NV)
color centers in diamond, whose structure and energy levels are summarized
in Fig.\ref{fig2}b and c. The ground state of NV centers has a spin
one with splitting $\omega_{\pm}/2\pi\simeq2.88\,$GHz between states
$m_{S}=0$ and $m_{S}=\pm1$ at zero magnetic field \cite{Jelezko2004}.
Each of the two $m_{S}=0$ to $m_{S}=\pm1$ transitions is further
splitted into three peaks separated by $2.2$~MHz due to the hyperfine
(HF) coupling to the $^{14}$N nuclear spin \cite{Felton2009}. In
the experiment a static magnetic field $B_{NV}=1.1$~mT \cite{BNV}
is applied along the $[1,1,1]$ crystallographic axis to lift the
degeneracy between states $m_{S}=\pm1$. Centers having their $N-V$
axis along $[1,1,1]$ (called in the following ensemble $I$) undergo
a different Zeeman shift from those along the three other $\left\langle 1,1,1\right\rangle $
axes (ensemble $III$), resulting in two different ESR frequencies
$\omega_{\mathrm{+I}}/2\pi=2.91$~GHz and $\omega_{\mathrm{+III}}/2\pi=2.89$~GHz
for the $m_{S}=0$ to $m_{S}=+1$ transition on which we will exclusively
focus in the following. 

The diamond crystal used is of the High-Pressure High-Temperature
(HPHT) $Ib$ type and has a NV center concentration of $\sim3$~ppm
and a residual nitrogen concentration of $\sim30$~ppm. In our setup,
it is glued on top of the ESR cavity $C$, a coplanar waveguide superconducting
resonator \cite{Frunzio2005} of quality factor $Q\sim10^{4}$ made
of a niobium thin-film sputtered on a silicon substrate. The spin
ensemble $S$ detected in the experiment consists of the $\sim10^{11}$
NV centers that lie within the mode volume of $C$, thus within a
few microns of the diamond surface. The cavity frequency $\omega_{c}(\Phi)$
can be tuned on a nanosecond timescale by a flux $\Phi$ applied through
the loop of a superconducting quantum interference device (SQUID)
inserted into $C$ \cite{TunableResonatorsPalacios,Sandberg,Kubo2010,Kubo2012}.
The superconducting qubit is a Cooper-pair box of the transmon type
\cite{transmon_th,transmon_exp} with resonance frequency $\omega_{ge}$
between its ground state $\left|g\right\rangle $ and excited state
$\left|e\right\rangle $. It is coupled to an additional resonator
$R$ which is nonlinear and used to read-out the qubit state. As explained
in details in Ref. \cite{Mallet2009}, this readout is performed in
a single shot by measuring the phase of a microwave pulse reflected
on $R$. This phase takes two different values depending on the qubit
state, and repeating $\sim10^{4}$ times the same experimental sequence
yields the qubit excited state probability $P_{e}$.

\begin{figure}[t]
\includegraphics[scale=0.9]{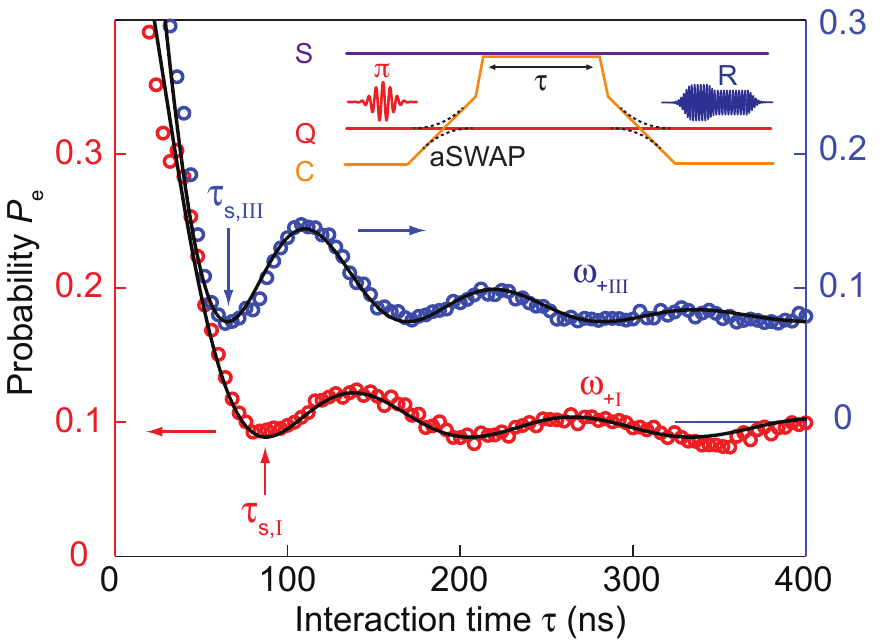}

\caption{Single photon transfer between qubit and spins \cite{KuboNVQubit2011}.
(inset): Pulse sequence used to excite the qubit in state $\left|e\right\rangle $
and transfer its excitation to $S$ via $C$ back and forth. (panel):
Swap oscillations for the two spin frequencies $\omega_{\mathrm{+I}}$
and $\omega_{\mathrm{+III}}$ . Swap times $\tau_{\mathrm{s,I}}$
and $\tau_{\mathrm{s,III}}$ are indicated by arrows. Note that the
two curves are shifted vertically for clarity with corresponding scales
on left (for ensemble $I$) and right axes (ensemble $III$). }

\label{fig3} 
\end{figure}

\section{Single photon storage and retrieval}

As reported in an earlier work \cite{KuboNVQubit2011} (see also \cite{Zhu2011}),
it is possible with this circuit to coherently exchange a single quantum
of excitation between the spin ensemble $S$ and the qubit $Q$ via
the tunable cavity $C$. To demonstrate that, the qubit is prepared
in $|e\rangle$; its excitation is transferred to the cavity by sweeping
adiabatically $\omega_{c}(\Phi)$ through $\omega_{ge}$, which is
then tuned suddently in resonance with the spins at $\omega_{K}$
(where $K=+I,+III$) for some interaction time $\tau$. The excitation
left in the cavity is finally transferred back into the qubit, which
is then read-out. As shown in Fig. \ref{fig3}, the resulting qubit
excited state probability $P_{e}(\tau)$ is found to oscillate, revealing
the conversion of a single microwave photon into an elementary collective
excitation of the spin ensemble. For well-defined interaction times
$\tau_{s,K}$ (see Fig. \ref{fig3}), the excitation in the qubit
is swapped into the spin ensemble \cite{KuboNVQubit2011}; at a later
time it is recovered in the qubit with a fidelity $\sim0.1$. 

A quantum-mechanical description of this experiment \cite{Diniz2011,Kurucz2011}
is useful in the discussion of the ESR results presented below. Each
of the $N_{K}$ effective spins of the ensemble at $\omega_{K}$ is
modelled as an effective harmonic oscillator with frequency $\omega_{j_{K}}$
and annihilation (resp. creation) operator $b_{j_{K}}$ (resp. $b_{j_{K}}^{\dagger}$),
an approximation valid in the low-excitation limit as is the case
throughout this article. The spin ensemble and cavity are then described
by Hamiltonians $\sum\hbar\omega_{j_{K}}b_{j_{K}}^{\dagger}b_{j_{K}}$
and $\hbar\omega_{c}(\Phi)a^{\dagger}a$, $a$ (resp. $a^{\dagger}$)
being the cavity annihilation (resp. creation) operator. The coupling
between the resonator and the spin ensemble is described by a Hamiltonian
$H_{K}=-i\hbar\sum g_{j_{K}}(b_{j_{K}}^{\dagger}a+h.c.)$, with $g_{j_{K}}$
the coupling constant of spin $j_{K}$ to the resonator. This Hamiltonian
can be rewritten as $H_{K}=-i\hbar g_{K}\left(b_{K}^{\dagger}a+h.c.\right)$
with $g_{K}=\left(\sum g_{j_{K}}^{2}\right)^{1/2}$ the spin ensemble-resonator
collective coupling constant and $b_{K}=(1/g_{K})\sum g_{\mathrm{j_{K}}}b_{\mathrm{j_{K}}}$
the annihilation operator of the collective spin excitation coupled
to the cavity. This superradiant mode has a spatial profile given
by the spatial dependence of the coefficients $g_{j_{K}}$, which
reproduces the profile of the magnetic field inside the cavity mode.
Note also that $b_{K}$ involves all the spins belonging to group
$K$, even if they have different frequencies due to slightly different
magnetic environment in the crystal, also including the three possible
states of the $^{14}N$ nuclear spin causing the hyperfine structure;
as a result this mode is coupled to $N_{K}-1$ dark modes that act
as a bath \cite{Kubo2012,Diniz2011,Kurucz2011,Sandner2011}. Using
these notations, one describes the oscillations shown in Fig. \ref{fig3}
as occurring between states $\left|1_{c},0_{K}\right\rangle $ and
$\left|0_{c},1_{K}\right\rangle $, where $\left|1_{c}\right\rangle =a^{\dagger}\left|0_{c}\right\rangle $
is the usual Fock state with $1$ photon in the cavity, and $\left|1_{K}\right\rangle =b_{K}^{\dagger}\left|0_{K}\right\rangle $
is the first excited state of the superradiant mode; damping of these
oscillations is due to inhomogeneous broadening, and can be interpreted
as damping of state $\left|1_{K}\right\rangle $ into the bath of
dark states \cite{Kubo2012}.

\section{ESR protocol and discussion}

\begin{figure}[t]
\includegraphics{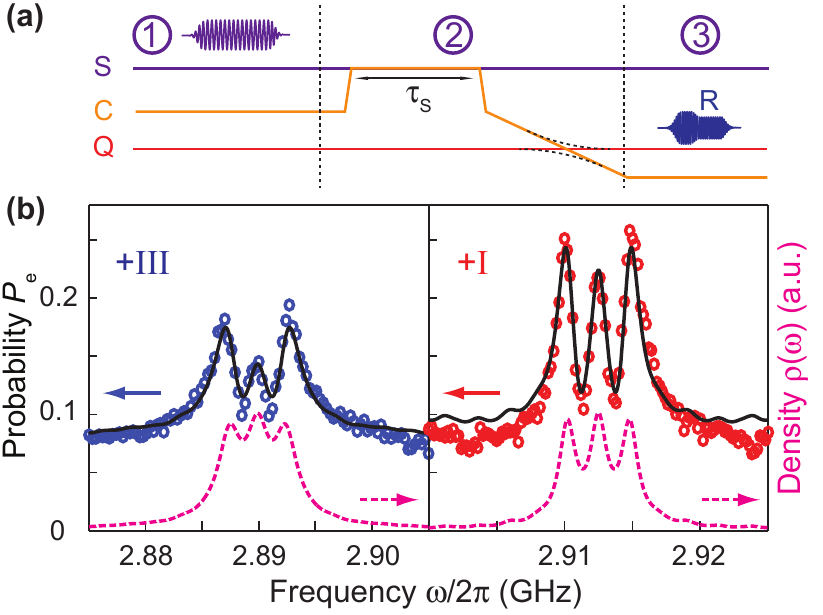}

\caption{\textbf{a}, Experimental pulse sequence used for qubit-detected ESR:
(1) The spins are first weakly excited by a $2\,\mu$s microwave pulse
with a frequency $\omega_{p}$; (2) the resulting spin excitation
is transferred to the cavity $C$ by a fast flux pulse which brings
$\omega_{c}$ in resonance with $\omega_{K}$ ($K=+I,+III)$ for a
swap time $\tau_{s,K}$, and then to the qubit $Q$ by an adiabatic
swap ($aSWAP$). (3) The qubit excited state probability $P_{e}$
is finally measured. \textbf{b,} Measured (open circles) and calculated
(solid line) $P_{e}(\omega_{p})$ for spin ensemble $+III$ (left)
and $+I$ (right). The spin density $\rho(\omega)$ used in the calculation
is shown as a dashed line.}

\label{fig4} 
\end{figure}

The ESR protocol is shown in Fig. \ref{fig4}. It consists in an experimental
sequence similar to the one used for single photon storage (see fig.
\ref{fig3}), but with the spin ensemble initially excited at several
photons level instead of the qubit prepared in its excited state.
More precisely, a low-power microwave pulse of duration $\Delta t=2\,\mu s$
and varying frequency $\omega_{p}$ is applied to the spins while
$\omega_{c}(\Phi)$ is far detuned; the resulting excitation is transferred
first into the cavity by tuning $\omega_{c}(\Phi)$ suddenly in resonance
with $\omega_{K}$ for the swap time $\tau_{s,K}$, then into the
qubit by an adiabatic swap interaction; the qubit state is finally
measured. Provided the average number of microwave photons emitted
by the spins into the ESR cavity stays much lower than $1$ to avoid
saturating the qubit, the resulting excited state probability $P_{e}(\omega_{p})$
is expected to reproduce the spin ensemble absorption spectrum. Experimental
results of Fig. \ref{fig4}b indeed display the characteristic HF
structure of NV centers consisting in three peaks separated by $2.2$~MHz
for both spin ensembles $+I$ and $+III$. This validates the concept
of electron spin resonance detected by a superconducting qubit. Note
that $P_{e}(\omega_{p})\ll1$ showing that qubit saturation is avoided
as wanted. 

We now discuss the sensitivity of this qubit-based ESR spectrometer.
The qubit state is detected in a single-shot with a fidelity of $\simeq0.7$
at the end of an experimental sequence that lasts typically $50\,\mu\mathrm{s}$,
yielding a $1\%$ precision on the probability $P_{e}$ in one second.
To translate this sensitivity in a magnetic moment unit, one needs
to know with what efficiency the excitation of the spin ensemble is
actually transferred to the qubit. The transfer of one microwave photon
from the cavity to the qubit is performed with an efficiency of order
unity (in our experiment it is around $0.7$ limited by losses in
the cavity and qubit), so the limiting factor is the efficiency of
the transfer of the spin ensemble excitation to the cavity during
their resonant interaction. At first sight one might think that since
the spin ensemble and cavity are in the strong coupling limit, one
excitation of the spin ensemble should also be converted into a microwave
photon with an efficiency of order $1$, similar to what happened
in the coherent oscillations shown in Fig. \ref{fig3}. This reasoning
is not correct here because the collective spin mode $b_{\omega_{p}}$
excited by the spectroscopy pulse does not necessarily match perfectly
the superradiant mode $b_{K}$. Indeed, although the spatial matching
of the two modes is excellent since the specroscopy pulse is applied
through the cavity, this is not the case for spectral matching: only
spins having a resonance frequency within the spectroscopy pulse bandwidth
$\delta/2\pi=150$~kHz around $\omega_{p}$ contribute to $b_{\omega_{p}}$,
whereas all spins within the hyperfine line (total width $\Delta/2\pi\sim5$~MHz)
contribute to the superradiant mode $b_{K}$. As a result, one expects
an overlap of order $\sqrt{\delta/\Delta}$ between the $b_{\omega_{p}}$
and $b_{K}$ modes, implying that there should be $\sim\Delta/\delta=2\text{0}$
times less excitations in the $b_{K}$ mode (and thus also in the
cavity mode after the swap interaction) than in the $b_{\omega_{p}}$
mode.

This argument can be made rigorous and quantitative. The collective
spin mode excited by the spectroscopy pulse is defined as $b_{\omega_{p}}=\left[\sum g_{j_{K}}\alpha_{\omega_{p}}(\omega_{j_{K}})b_{j_{K}}\right]/\sqrt{\sum g_{j_{K}}^{2}\left|\alpha_{\omega_{p}}(\omega_{j_{K}})\right|^{2}}$,
with $\alpha_{\omega_{p}}(\omega)$ the pulse Fourier transform. The
quantity of interest is then the correlation function $\beta(\omega_{p},\tau_{s})=\left\langle a(\tau_{s})b_{\omega_{p}}^{\dagger}(0)\right\rangle $
giving the probability amplitude for an excitation created in $b_{\omega_{p}}$
by the spectroscopy pulse to be transferred into a photon inside the
cavity mode after an interaction time $\tau_{s}$. This function can
be computed numerically given a certain spin distribution $\rho(\omega)$
using the formulas derived in the Appendix. In our experiment, the
linewidth of each hyperfine peak $w_{\text{+}I}/2\pi=1.6$~MHz and
$w_{+III}/2\pi=2.4$~MHz, and coupling constants $g_{+I}/2\pi=2.9$~MHz
and $g_{\text{+}III}/2\pi=3.8$~MHz have been determined from other
measurements \cite{KuboNVQubit2011}, so that a direct comparison
with theory without any adjustable parameter is possible as shown
in Fig. \ref{fig4}. The agreement is quantitative (note that we have
also included in the distributions $\rho(\omega)$ additional ESR
frequencies caused by the hyperfine interaction of the NV center with
neighboring $^{13}C$ nuclei with the $1.1\%$ natural abundance as
expected). From this calculation, we deduce that the average excitation
of the spin ensemble at resonance in the data shown in Fig. \ref{fig4}
is $\sim15$, in agreement with the qualitative argument presented
above. In the present state of the experiment, the qubit-based ESR
spectrometer therefore measures the spectrum of an ensemble of $10^{11}$
NV centers at an excitation level of order $15\,\mu_{B}$, in a one
minute integration time. 

Thanks to this very low excitation level, the experimental sequence
can be safely repeated at $20$~kHz, despite the NV centers energy
relaxation time reaching minutes at $30$~mK \cite{Majer2011,Budker2011}.
More precisely, two factors contribute to make this experiment possible:
1) at the end of each experimental sequence the excitation of the
$b_{\omega_{p}}$ mode quickly decays into the bath of dark modes,
allowing the next experimental sequence to start with $b_{\omega_{p}}$
in its ground state and thus keeping the average number of excitations
transfered to the qubit well below $1$ as needed to avoid saturation,
2) the low excitation rate ensures on the other hand that the ensemble
of $10^{11}$ spins stays far from saturation even after repeating
the sequence for hours. 

We finally note that our calculation reproduces a puzzling feature
of the data that was not discussed yet: the middle peak of the $P_{e}(\omega_{p})$
curve has a lower amplitude than the two other peaks, both for the
$+I$ and the $+III$ curves as seen in Fig. \ref{fig4}, although
the spin density $\rho(\omega)$ used in the calculation is a simple
sum of three Lorentzians with the same amplitude. Our ESR protocol
thus appears to slightly distort the absorption spectrum. This phenomenon
originates from the $\omega_{p}$ dependence of the energy transfer
efficiency from the spins into the cavity, caused by the fact that
$g_{K}\approx\Delta$ in our sample. It could probably be corrected
in future experiments either by increasing $g_{K}$ or by transferring
the spin excitation to the cavity with an adiabatic passage.

Besides detecting a large ensemble of $N=10^{11}$ electronic spins
at near single-excitation level, it is interesting to discuss what
is the minimal number of spins $N_{min}$ that could be detected with
a similar experimental protocol in order to compare it to the sensitivity
of existing conventional spectrometers. For that we will change perspective
in the following discussion, and assume that the spins being measured
can actually be excited at saturation. We suppose that the $N$ spins,
of inhomogeneous linewidth $\Delta$, have been excited by a hard
$\pi/2$ pulse. In the weak coupling limit $g\sqrt{N}\ll\kappa\ll\Delta$
($\kappa=\omega_{c}/Q$ being the cavity damping rate), the spins
emit in the cavity $\bar{n}=g^{2}N^{2}/(4\Delta^{2})$ photons (see
Appendix B). Taking a conservative estimate for the minimal average
excitation that can be detected by a superconducting qubit within
one second to be $0.05$, a spin-cavity coupling constant $g/2\pi=10$~Hz,
one obtains $N_{min}=10^{5}$ spins$/\sqrt{\mathrm{Hz}}$ for a $0.1$~mT
linewidth corresponding to $\Delta/2\pi=2.8$~MHz. This figure is
five orders of magnitude better than a commercial spectrometer at
$300$~K, and two orders of magnitude better than the record sensitivity
of $10^{6}$ spins$/\sqrt{Hz}$ for a $0.01$~mT linewidth that was
recently reported with a surface loop-gap resonator operated at $10$~K
\cite{Twig2011} and a coplanar waveguide resonator at $4$~K \cite{Schuster2012}.
Note however that in order to operate such a qubit-based spectrometer
in practice, one would need 1) to use a repetition rate around $10$~kHz
which requires a reasonably short spin-lattive relaxation time or
some way to repump rapidly the spins into their ground state (optically
\cite{Yang2009} or electrically for instance) and 2) to find a qubit
design that withstands large magnetic fields usually needed for ESR.
Interesting alternative possibilities could be to physically separate
the qubit-detector from the spins, which would allow more easily to
apply large magnetic fields to the spins without perturbing the qubit,
and to continuously monitor the qubit state with a parametric amplifier
as demonstrated in recent experiments \cite{Vijay2011} instead of
pulsing the qubit state detection as done here. 

In conclusion we have discussed a new type of ESR spectrometer in
which the signal coming from the spins is detected by a superconducting
qubit acting as a single-microwave-photon detector. We have implemented
this idea on an ensemble of $\sim10^{11}$ NV centers coupled to a
transmon qubit, measuring their absorption spectrum at an excitation
level of $\sim15\,\mu_{B}$, with a well-resolved hyperfine structure.
Estimates indicate that this spectrometer would be able to detect
$10^{5}$ spins$/\sqrt{\mathrm{Hz}}$ with a $0.1$~mT linewidth,
a gain of two orders of magnitude in sensitivity compared to the best
reported values for conventional spectrometers. Our work thus demonstrates
the potential of superconducting circuits for electron spin resonance
spectroscopy.

\section*{Appendix A}

We now explain in more detail how the theory curves in figure 3 are
calculated. As explained in the main text, the energy transfer efficiency
from the resonator to the qubit is of order unity and can be well
modeled by an ideal adiabatic passage. In this context the quantity
of interest is the resonator population after the interaction with
the spins. The calculations are performed in the Holstein-Primakoff
approximation, in which the spins and the resonator are described
by harmonic oscillators. The total system Hamiltonian is $H/\hbar=\omega_{\mathrm{c}}(\Phi)a^{\dagger}a+\sum\omega_{\mathrm{j}}b_{\mathrm{j}}^{\dagger}b_{\mathrm{j}}+\sum ig_{\mathrm{j}}(b_{\mathrm{j}}^{\dagger}a-b_{\mathrm{j}}a^{\dagger})$,
$g_{\mathrm{j}}$ being the coupling constant of spin $j$ with the
resonator. We need to calculate the probability that the excitation
created at $t=0$ in the spins to be transferred to the cavity after
a time $t$, this probability is the square modulus of $\left\langle 0\right|a(t)b_{\omega_{p}}^{\dagger}\left|0\right\rangle $.
The spins excitation is created by a microwave pulse of central frequency
$\omega_{p}$ with a pulse envelope in frequency described by $\alpha_{\omega_{p}}(\omega)=\alpha(\omega-\omega_{p})$,
a typical envelope is a Lorentzian function with FWHM $\delta$. We
can define an operator $b_{\omega_{p}}^{\dagger}$ that describes
the excitation induced by this pulse as 
\begin{equation}
b_{\omega_{p}}^{\dagger}=\frac{1}{\sqrt{{\displaystyle {\sum_{j}|\alpha_{\omega_{p}}(\omega_{j})|^{2}g_{j}^{2}}}}}\quad\sum_{k}\alpha_{\omega_{p}}(\omega_{k})g_{k}b_{k}^{\dagger}\,,\label{eq:1}
\end{equation}
 this comes simply from the standard atom-field interaction for a
classical light source such as the one used in the experiment. As
shown in \cite{Diniz2011} the quantity $\left\langle 0\right|a(t)b_{\omega_{p}}^{\dagger}\left|0\right\rangle $
can be calculated by considering an effective non-Hermitian Hamiltonian
\begin{equation}
H_{eff}/\hbar=\left(\begin{array}{cccc}
\tilde{\omega}_{0} & ig_{1} & ig_{2} & \ldots\\
-ig_{1} & \tilde{\omega}_{1}\\
-ig_{2} &  & \tilde{\omega}_{2}\\
\vdots &  &  & \ddots
\end{array}\right)\,.
\end{equation}
 with complex angular frequencies $\tilde{\omega}_{0}=\omega_{c}(\Phi)-i\kappa/2$
and $\tilde{\omega}_{k}=\omega_{k}-i\gamma_{0}/2$ ; here, $\gamma_{0}$
is the spontaneous emission rate of each spin (that we take here to
be zero since NV centers at low temperature have negligible energy
relaxation). Indeed, introducing the vector $X(t)$ of coordinates
$\left[\left\langle a(t)a^{\dagger}(0)\right\rangle ,...,\left\langle b_{j}(t)a^{\dagger}(0)\right\rangle ,...\right]$
it can be shown that $dX/dt=-(i/\hbar)H_{eff}X$. The formal solution
to this equation is then

\begin{equation}
X(t)=\mathcal{L}^{-1}[(s+iH_{eff}/\hbar)^{-1}X(0)]\,,\label{eq:2}
\end{equation}

which gives $\left\langle 0\right|a(t)b_{\omega_{p}}^{\dagger}\left|0\right\rangle =x_{G}{}^{\dagger}\cdot X(t)=\mathcal{L}^{-1}\left[t_{\omega_{p}}(s)\right]$
with $x_{G}=(1,0,0,...)$ and $\mathcal{L}[f(s)]=\int e^{-st}f(t)dt$
($s$ being a complex number). The initial condition $X(0)$ is the
one produced by $b_{\omega_{p}}^{\dagger}$ given in Eq.\ref{eq:1},
thus 
\begin{equation}
\begin{split}t_{\omega_{p}}(-i\omega) & =\frac{\sum_{k}\alpha_{\omega_{p}}(\omega_{k})g_{k}}{\sqrt{{\sum_{j}|\alpha_{\omega_{p}}(\omega_{j})|^{2}g_{j}^{2}}}}\;\;[(s+iH_{eff})^{-1}]_{0,k}\\
 & =\frac{\sum_{k}\alpha_{\omega_{p}}(\omega_{k})g_{k}}{\sqrt{{\sum_{j}|\alpha_{\omega_{p}}(\omega_{j})|^{2}g_{j}^{2}}}}\;\;\left[\frac{g_{k}\; t_{1}(-i\omega)}{i\gamma_{0}+(\omega-\omega_{p})}\right]\\
 & =\frac{t_{1}(-i\omega)}{i\gamma_{0}+(\omega-\omega_{p})}\frac{\sum_{k}\alpha_{\omega_{p}}(\omega_{k})g_{k}^{2}}{\sqrt{\sum_{j}|\alpha_{\omega_{p}}(\omega_{j})|^{2}g_{j}^{2}}}\;,
\end{split}
\end{equation}
 where $t_{1}(-i\omega)=i/\left[\omega-\omega_{0}+i\kappa/2-W(\omega)\right]$
with $W(\omega)=\sum_{j}g_{j}^{2}/\left[\omega-\omega_{j}+i\gamma_{0}/2\right]$.
Note that we evaluated $t_{\omega_{p}}(s)$ for $s=-i\omega$, this
is sufficient to perform the Laplace transform inversion as there
are no singularities in the imaginary axis of $t_{\omega_{p}}$. We
define the spin density $\rho(\omega)$ encompassing the coupling
strength, which is possibly different for each spin, as $\rho(\omega)=\sum_{j}\frac{g_{j}^{2}}{g_{K}^{2}}\;\delta(\omega-\omega_{j})$.
Using this definition in the equation above we have 
\begin{equation}
\begin{split}t_{\omega_{p}}=\frac{g_{K}\; t_{1}(-i\omega)}{i\gamma_{0}+(\omega-\omega_{p})}\frac{(\alpha\ast\rho)(\omega_{p})}{\sqrt{(|\alpha|^{2}\ast\rho)(\omega_{p})}}\,.\end{split}
\end{equation}

The spectral width of the microwave pulse is, in our case, much smaller
than any scale that characterizes our distribution $\rho(\omega)$.
This allows the rewriting of the convolution above as 
\begin{equation}
\begin{split}\frac{(\alpha\ast\rho)(\omega_{p})}{\sqrt{(|\alpha|^{2}\ast\rho)(\omega_{p})}}=A\sqrt{\rho(\omega_{p})}\;,\end{split}
\end{equation}
 where the constant $A=\frac{\int\alpha(\omega)d\omega}{\sqrt{\int|\alpha(\omega)|^{2}d\omega}}$
is purely characterized by the pulse envelope with no dependence on
$\omega_{p}$, yielding for example $A=\sqrt{\delta}\sqrt{\pi/2}$
for a Lorentzian envelope. This means that if we consider that the
spins are distributed at a typical range $\Delta$ the equation above
gives a rigorous justification of the rule of thumb that says that
the efficiency of the spins-resonator transfer is given by the overlap
$\sqrt{\delta/\Delta}$.



Finally to generate the theoretical curve in Fig. 3, we perform a
numerical inversion of the Laplace transform for each $\omega_{p}$
and take $|\left\langle 0\right|a(t)b_{\omega_{p}}^{\dagger}\left|0\right\rangle |^{2}$
at $t=\tau_{S,III}$ or $t=\tau_{S,I}$.

\section*{Appendix B}

We now explicit the calculation of the sensitivity of our qubit-based
ESR spectrometer in the weak coupling limit $g_{K}\ll\kappa\ll\Delta$
. The Hamiltonian coupling the spins to the cavity field is $H=\hbar g(S_{-}a^{\dagger}+h.c.)$,
where $S_{-}=\sum\sigma_{i,-}$, $\sigma_{i,-}$ being the lowering
operator of spin $i$. In the absence of driving field, the equation
for the intra-cavity mean field is then easily obtained as 

\[
\frac{d<a>}{dt}=-\frac{\kappa}{2}<a>-ig<S_{-}>.
\]

Right after a $\pi/2$ pulse on the spins, $|<S_{-}>|=N/2$. Neglecting
the back-action of the cavity field on the spins (which is justified
in the weak coupling limit), we simply get that $<S_{-}>=(N/2)e^{-\Delta t}$.
From that one shows that $<a>(t)=-igN(e^{-\kappa t/2}-e^{-\Delta t})/[\kappa-2\Delta]$
which in the limit $\kappa\ll\Delta$ yields a maximum photon number
in the cavity of $\bar{n}=g^{2}N^{2}/(4\Delta^{2})$.

\textbf{Acknowledgements} We acknowledge useful discussions with B.
Julsgaard, K. Moelmer, J. Morton, A. Briggs, and within the Quantronics
group, and technical support from P. Sénat, P.-F. Orfila, T. David,
J.-C. Tack, P. Pari, P. Forget, M. de Combarieu. We acknowledge support
from European project Solid, ANR projects Masquelspec and QINVC, C'Nano,
Capes, and Fondation Nanosciences de Grenoble.


\begin{thebibliography}{References}
\bibitem{Kubo2010} Y. Kubo et al., Phys. Rev. Lett. \textbf{105},
140502 (2010).

\bibitem{Schuster2010} D.I. Schuster, et al., Phys. Rev. Lett. \textbf{105},
140501 (2010).

\bibitem{Majer2011} R. Amsuss, et al., Phys. Rev. Lett. \textbf{107},
060502 (2011).

\bibitem{Bushev2011} P. Bushev et al., Phys. Rev. B \textbf{84},
060501 (2011).

\bibitem{Schuster2012} H. Malissa, D. I. Schuster, A.M. Tyryshkin,
A.A. Houck, and S.A. Lyon, arxiv:1202.6305.

\bibitem{Chen2011} Y.-F. Chen et al., Phys. Rev. Lett. \textbf{107},
217401 (2011).

\bibitem{WendinShumeiko2007} G. Wendin and V.S. Shumeiko, Low Temperature
Physics \textbf{9}, 724 (2007).

\bibitem{ESRnote} Note that in a real ESR spectrometer, the FID signal
would in fact be difficult to detect using this protocol because of
ringing of the cavity and of the detectors dead-time. The spin absorption
spectrum would therefore rather be measured by CW-ESR rather than
by this FID-detected protocol. We nevertheless discuss it for pedagogical
reasons, because it is closest to our qubit-detected ESR experiment
in which these technical issues are irrelevant.

\bibitem{KuboNVQubit2011} Y. Kubo et al., Phys. Rev. Lett. \textbf{107},
220501 (2011).

\bibitem{Jelezko2004} F. Jelezko, T. Gaebel, I. Popa, A. Gruber,
and J. Wrachtrup,\textit{ }Phys. Rev. Lett. \textbf{92}, 076401 (2004).

\bibitem{Felton2009} S. Felton, \textit{et al.}, Phys. Rev. B \textbf{79},
075203 (2009).

\bibitem{BNV} The experiment was performed in the same magnetic field
as in \cite{KuboNVQubit2011} where the reported value was mistakenly
mentioned to be $B_{NV}=1.4$~mT. 

\bibitem{Frunzio2005} L. Frunzio, A. Wallraff, D. I. Schuster, J.
Majer, and R. J. Schoelkopf, IEEE Trans. on Applied Superconductivity
\textbf{15}, 860 (2005).

\bibitem{TunableResonatorsPalacios} A. Palacios-Laloy \emph{et al.},
J. Low Temp. Phys. \textbf{151}, 1034 (2008).

\bibitem{Sandberg} M. Sandberg, \emph{et al.}, Appl. Phys. Lett.
\textbf{92}, 203501 (2008).

\bibitem{Kubo2012} Y. Kubo, \textit{et al.}, Phys. Rev. A \textbf{85},
012333 (2012).

\bibitem{transmon_th} J. Koch \textit{et al.}, Phys. Rev. A \textbf{76,}
042319 (2007).

\bibitem{transmon_exp} J. A. Schreier \textit{et al., }Phys. Rev.
B \textbf{77,} 180502 (2008).

\bibitem{Mallet2009}  F. Mallet, \textit{et al.}, Nature Phys. \textbf{5},
791 (2009).

\bibitem{Budker2011} A. Jarmola, V. M. Acosta, K. Jensen, S. Chemerisov,
D. Budker, arxiv:1112.5936

\bibitem{Zhu2011} X. Zhu \textit{et al.}, Nature \textbf{478}, 221
(2011).

\bibitem{Diniz2011} I. Diniz\textit{ et al.}, Phys. Rev. A \textbf{84},
063810 (2011).

\bibitem{Kurucz2011} Z. Kurucz, J.H. Wesenberg and K. Molmer, Phys.
Rev. A \textbf{83}, 053852 (2011).

\bibitem{Sandner2011} K. Sandner \textit{et al.}, arxiv:1112.4767
(2011).

\bibitem{Yang2009} A. Yang \textit{et al.}, PRL \textbf{102}, 257401
(2009).

\bibitem{Twig2011} Y. Twig, E. Dikarov, W.D. Hutchison, and A. Blank,
Rev. of Sci. Instr. \textbf{82}, 076105 (2011).

\bibitem{Vijay2011} R. Vijay \textit{et al.}, Phys. Rev. Lett. \textbf{106},
110502 (2011).\end{thebibliography}
\end{document}